\documentclass[12pt,reqno]{article}
\usepackage{amsmath}

\allowdisplaybreaks

\renewcommand{\i}{\mathrm{i}}
\newcommand{\e}{\mathrm{e}}
\newcommand{\al}{\alpha}
\newcommand{\ad}{{\dot{\alpha}}}
\newcommand{\be}{\beta}
\newcommand{\bd}{{\dot{\beta}}}
\newcommand{\ga}{\gamma}
\newcommand{\de}{\delta}
\newcommand{\ep}{\varepsilon}
\newcommand{\bpsi}{\bar{\psi}}
\newcommand{\si}{\sigma}
\newcommand{\bsi}{\bar{\sigma}}
\newcommand{\la}{\lambda}
\newcommand{\bla}{\bar{\lambda}}
\newcommand{\cD}{\mathcal{D}}
\newcommand{\cF}{\mathcal{F}}
\newcommand{\half}{\tfrac{1}{2}}
\newcommand{\quart}{\tfrac{1}{4}}
\newcommand{\tab}{\quad\,}
\newcommand{\p}{\partial}

\newcommand{\bD}{\bar{D}}
\newcommand{\bcD}{\bar{\mathcal{D}}}
\newcommand{\Dz}{\Delta_z}
\newcommand{\com}[2]{[\,#1\, ,\,#2\,]}
\newcommand{\aco}[2]{\{#1\, ,\,#2\}}
\newcommand{\frc}[2]{\frac{\raisebox{-2pt}{$#1$}}{#2}}
\DeclareMathOperator{\im}{Im}
\DeclareMathOperator{\re}{Re}

\DeclareSymbolFont{AMSa}{U}{msa}{m}{n}
\DeclareSymbolFont{AMSb}{U}{msb}{m}{n}
\DeclareMathSymbol{\square}{\mathord}{AMSa}{"03}
\DeclareMathSymbol{\fieldR}{\mathalpha}{AMSb}{"52}

\begin{document} 

\begin{flushright} \small
 YITP--SB--00--82 \\ hep-th/0012096
\end{flushright}
\medskip

\begin{center}
 {\large\bfseries Nonlinear Vector-Tensor Multiplets Revisited}
 \\[5mm]
 Ulrich Theis
 \\[2mm]
 {\small\slshape
 C.N.~Yang Institute for Theoretical Physics \\
 State University of New York at Stony Brook \\
 Stony Brook, NY 11794-3840, USA \\[2pt]
 theis@insti.physics.sunysb.edu}
\end{center}
\vspace{5mm}

\hrule\bigskip

\centerline{\bfseries Abstract} \medskip
We give an off-shell formulation of the $N=2$ supersymmetric new
nonlinear vector-tensor multiplet. Interactions arise in this model as a
consequence of gauging the central charge of the supersymmetry algebra,
which in contrast to previous models with local central charge is
achieved without a coupling to a vector multiplet. Furthermore, we
present a new action formula that follows from coupling the $N=2$
linear multiplet to the vector-tensor multiplet.
\bigskip

\hrule\vspace{.3in}


In their study of four-dimensional $N=2$ supersymmetric vacua of
heterotic string theory de~Wit et al.\ \cite{dWKLL} found that the
dilaton and the antisymmetric tensor reside in an $N=2$ vector-tensor
multiplet. Although on-shell a vector-tensor (VT) multiplet is dual to
an abelian vector multiplet, string theory may prefer the off-shell
realization of the former. This is why possible interactions of the VT
multiplet have been extensively studied in the last years. Among these,
which include a variety of couplings to additional vector multiplets
\cite{CdWFKST}--\cite{DT} and supergravity couplings \cite{CdWFKST2},
there are self-interactions.

A self-interacting VT multiplet with nonlinear supersymmetry
transformations had been constructed for the first time by Claus et
al.\ in \cite{CdWFKST}. Only recently was it realized that there exists
a second self-interacting version of the multiplet: in \cite{T} we
obtained what we shall call the new nonlinear vector-tensor (NLVT)
multiplet in the following by gauging its central charge
transformations. Promoting the global bosonic symmetry associated with
the central charge of the VT multiplet to a local symmetry requires the
presence of a vector potential, which in previous works was taken to sit
in an additional abelian vector multiplet \cite{CdWFKST,CdWFT,DIKST,DT}.
In \cite{T} on the other hand it was shown that the VT multiplet's own
vector field can serve as the necessary gauge connection, as well. The
resulting new NLVT multiplet features nonpolynomial but local
interactions. They were constructed by means of the Noether method as a
deformation of the free action and its symmetries, with the
supersymmetry algebra being realized only on-shell.

In the present paper we derive an off-shell formulation of the new NLVT
multiplet from a suitable set of constraints on the field strengths
appearing in the $N=2$ supersymmetry algebra with a gauged central
charge. The Bianchi identities imply that as a consequence of these
constraints all field strengths can be expressed in terms of a single
real scalar superfield, the components of which comprise the covariant
fields of the VT multiplet. This superfield in turn is subject to
constraints that give rise to the nonpolynomial dependence of the field
strengths on the vector. In order to obtain an invariant action, we
adapt a well-known construction principle that employs the properties of
(composite) linear multiplets \cite{BS,dWvHvP}, which as we show can be
realized as representations of our new vector-tensor supersymmetry
algebra by modifying their defining superfield constraints.

We should remark at this point that the term ``new nonlinear
vector-tensor multiplet'' has already been used in \cite{IS} to describe
a model different from ours. While this model was derived from a
deformation of the superfield constraints of the VT multiplet, there is
no realization in terms of the usual field content that includes an
antisymmetric tensor gauge potential, and a nontrivial action can only
be obtained by means of a dualization into a vector multiplet. Since no
such problems occur in our model, we thought it appropriate to call it
the ``new NLVT multiplet'' in order to distinguish it from the other
proper nonlinear VT multiplet of ref.\ \cite{CdWFKST}. We hope this does
not lead to confusion.
\bigskip

\textbf{Linear VT multiplet and Possible Deformations} 
\medskip

We consider rigid $N=2$ supersymmetry with a real central charge. The
supersymmetry algebra is spanned by supercovariant derivatives
 \begin{equation}
  D_A \in \big\{ \p_\mu,\, D_\al^i,\, \bD_{\ad i},\, \Dz \big\}\ .
 \end{equation}
We shall not distinguish properly between superfields and components,
so the $D_A$ may either be read as differential operators in superspace
or as generators of symmetry transformations of component fields. The
$D_\al^i$ and $\bD_{\ad i}$ are two-component Weyl spinors (greek
indices) as well as SU(2) doublets (small latin indices), where
$\bD_{\ad i}=\overline{D_\al^i}$. $\Dz$ denotes the generator of global
central charge transformations.

The graded commutator of two supercovariant derivatives involves
torsion,
 \begin{equation} \label{free_alg}
  \com{D_A}{D_B} = - T_{AB}{}^C D_C\ ,
 \end{equation}
where the nonvanishing components are\footnote{Our conventions are the
following: $\eta_{00}=+1$, $\si^\mu=(1,\vec{\si})$, $\si^{\mu\nu}=\half
\si^{[\mu}\bsi^{\nu]}$, $\ep^{12}=-\ep_{12}=1$. (Anti-) symmetrization
is defined such that $S_{\mu\nu}=S_{(\mu\nu)}+S_{[\mu\nu]}$. A tilde
denotes Hodge-dualization, $\tilde{F}^{\mu\nu}=\half\ep^{\mu\nu\rho\si}
F_{\rho\si}$.}
 \begin{equation} \label{torsion}
  T_\al^i{}^{}_{\,\ad j}{}^\mu = \i \de^i_j \si^\mu_{\al\ad}\ ,\quad
  T_\al^i{}_\be^j{}^z = - \i \ep_{\al\be}\, \ep^{ij}\ ,\quad T^{}_{\ad
  i}{}_{\,\bd j}{}^z = \i \ep_{\ad\bd} \ep_{ij}\ .
 \end{equation}

The VT multiplet \cite{SSW} is a representation of the algebra
$\eqref{free_alg}$ with a nontrivial central charge. The bosonic field
content consists of two real scalars $\phi$ and $U$, the latter being
an auxiliary field, a 1-form with components $A_\mu$ and a 2-form
$B_{\mu\nu}$. The fermions $\psi_\al^i$, $\bpsi_{\ad i}$ form SU(2)
doublets. With $F_{\mu\nu}=2\p_{[\mu}A_{\nu]}$ and $H^\mu=\half
\ep^{\mu\nu\rho\si}\p_\nu B_{\rho\si}$ we denote the field strength
of $A_\mu$ and the Hodge-dual field strength of $B_{\mu\nu}$
respectively.

The free action with Lagrangian
 \begin{equation}
  \mathcal{L}^{(0)} = \frc{1}{2} \p^\mu \phi\, \p_\mu \phi - \frc{1}{4}
  F^{\mu\nu} F_{\mu\nu} - \frc{1}{2} H^\mu H_\mu - \i \psi^i \si^\mu
  \overset{\leftrightarrow}{\p_\mu} \bpsi_i + \frc{1}{2} U^2
 \end{equation}
is, among others, invariant under gauge transformations of the
potentials,
 \begin{equation} \label{free_del}
  \de_\mathrm{gauge} A_\mu = \p_\mu C\ ,\quad \de_\mathrm{gauge}
  B_{\mu\nu} = \p_\mu C_\nu - \p_\nu C_\mu\ ,
 \end{equation}
linear supersymmetry transformations generated by
 \begin{align}
  D_\al^i \phi & = \psi_\al^i & D_\al^i \psi^{\be j} & = \half
	\ep^{ij} (F_{\mu\nu} \si^{\mu\nu} - \i U)_\al{}^\be \notag
	\\[2pt]
  D_\al^i A_\mu & = \i (\si_\mu \bpsi^i)_\al & D_\al^i \bpsi_\ad^j
	& = \half \ep^{ij} \si^\mu_{\al\ad} (H_\mu + \i \p_\mu \phi)
	\notag \\[2pt]
  D_\al^i B_{\mu\nu} & = 2 (\si_{\mu\nu} \psi^i)_\al & D_\al^i U & =
	(\si^\mu \p_\mu \bpsi^i)_\al\ ,
 \end{align}
and global central charge transformations generated by
 \begin{align}
  \Dz \phi & = U & \Dz \psi_\al^i & = (\si^\mu \p_\mu \bpsi^i)_\al
	\notag \\[2pt]
  \Dz A_\mu & = H_\mu & \Dz \bpsi_\ad^i & = - (\p_\mu \psi^i \si^\mu
	)_\ad \notag \\[2pt]
  \Dz B_{\mu\nu} & = \tilde{F}_{\mu\nu} & \Dz U & = \square \phi\ .
 \end{align}
The central charge transformations of the gauge potentials leave the
corresponding field strengths invariant on-shell (although the symmetry
is nontrivial, i.e.\ it does not reduce to a gauge transformation
on-shell) and are examples of the ``hidden'' symmetries considered in
\cite{B1}.

Self-interactions of the VT multiplet may be introduced by deforming the
free Lagrangian, i.e. by adding terms $\mathcal{L}^{(k)}$, $k=1,2,\dots$
to $\mathcal{L}^{(0)}$, where $\mathcal{L}^{(k)}$ is of order $k$ in
some continuous deformation parameters $g$. In order to maintain the
symmetries, it is necessary also to deform the corresponding field
transformations, so the generators decompose into pieces of definite
order in the $g$'s, as well, with $\de_\mathrm{gauge}$, $D_\al^i$ and
$\Dz$ as above being the $0$th-order contributions\footnote{A review of
deformation theory which focuses on $N=2$ supersymmetry is contained
in \cite{B2}.}.

To first order in $g$, invariance of the deformed action under the
deformed symmetries requires $\de^{(0)}\mathcal{L}^{(1)}\approx 0$
modulo total derivatives, where the relation $\approx$ denotes on-shell
(with respect to the linearized equations of motion) equality and
$\de^{(0)}$ is a $0$th-order symmetry generator. It can be shown by
writing down the general ansatz for the first-order deformation
$\mathcal{L}^{(1)}$ and imposing the above condition that there are
precisely two consistent SU(2) invariant deformations of the free VT
multiplet that involve vertices of mass dimension five, namely
 \begin{align}
  \mathcal{L}^{(1)} & = g_1 \big[ A_\mu F^{\mu\nu} H_\nu - \half \phi
	\tilde{F}^{\mu\nu} F_{\mu\nu} - \i F_{\mu\nu} (\psi^i
	\si^{\mu\nu} \psi_i - \bpsi^i \bsi^{\mu\nu} \bpsi_i) \big]
	\notag \\[2pt]
  & \tab + g_2 \big[ A_\mu \tilde{F}^{\mu\nu} H_\nu + \half \phi
	F^{\mu\nu} F_{\mu\nu} - \phi H^\mu H_\mu + F_{\mu\nu} (\psi^i
	\si^{\mu\nu} \psi_i + \bpsi^i \bsi^{\mu\nu} \bpsi_i) \notag \\
  & \mspace{65mu} + 2 H_\mu\, \psi^i \si^\mu \bpsi_i \big]\ .
 \end{align}
$\mathcal{L}^{(1)}$ is unique modulo total derivatives and on-shell
trivial terms that can be removed by means of field redefinitions.

The first term in each deformation is of the form $g_a A_\mu J^\mu_a$,
where the $J^\mu_a$ are conserved currents, and yields deformations of
the gauge transformations \eqref{free_del}. The current $J^\mu_1=
F^{\mu\nu}H_\nu$ corresponds to the global symmetry of $\mathcal{L}^{
(0)}$ generated by the central charge $\Dz$. The current $J^\mu_2=
\tilde{F}^{\mu\nu}H_\nu$ is trivial, $J_2^\mu\approx\p_\nu S^{[\mu\nu]}$
with $S^{[\mu\nu]}=\ep^{\mu\nu\rho\si}\!A_\rho H_\si$, but it still
deforms the gauge transformations (actually, only the one of $B_{\mu
\nu}$) since $S^{[\mu\nu]}$ is not gauge invariant\footnote{$J^\mu_2$
is trivial in the characteristic cohomology, but not in the invariant
characteristic cohomology \cite{HKS}.}. These two terms are examples
of so-called Henneaux-Knaepen and Chapline-Manton vertices respectively,
cf.\ \cite{HK,B3}.

The deformations can be completed separately to local actions invariant
to all orders in the coupling constants $g_a$, which in both cases
requires an infinite number of $\mathcal{L}^{(k)}$. But whereas the full
deformation with parameter $g_2$ is nonpolynomial only in the scalar
field $\phi$, the deformation with $g_1$ turns out to be nonpolynomial
also in the gauge field $A_\mu$. The same applies to the gauge
transformations in the respective deformations. The $g_2$-deformation
was constructed first by Claus et al.\ in \cite{CdWFKST} by means of the
superconformal multiplet calculus, and later in \cite{DK,IS} using
harmonic superspace techniques. The $g_1$-deformation was discovered
recently in \cite{T}, but has not been formulated off-shell yet.

In order to compare the formulation of the old NLVT multiplet with that
of the new one to be constructed in the following, we briefly
recapitulate the approach of \cite{IS}: The free VT multiplet is
given in terms of a field $\phi$ that satisfies the constraints
\cite{HOW}
 \begin{equation}
  D^{(i} D^{j)} \phi = 0\ ,\quad D_\al^{(i} \bD_\ad^{j)} \phi = 0\ .
 \end{equation}
One then considers deformations of these constraints that preserve
the commutation relations \eqref{free_alg}. The general ansatz
compatible with the Lorentz and SU(2) symmetry properties is given by
 \begin{equation}
  D^{(i} D^{j)} \phi = F(\phi)\, D^i \phi\, D^j \phi + G(\phi)\,
  \bD^i \phi\, \bD^j \phi\ ,\quad D_\al^{(i} \bD_\ad^{j)} \phi = 0
 \end{equation}
(the latter expression can always be made to vanish by a field
redefinition), where $F$ and $G$ are arbitrary functions of $\phi$.
The supersymmetry algebra imposes consistency conditions that constrain
the coefficient functions. Necessary, but not sufficient, are the
relations
 \begin{equation} \label{dF=FF}
  \frc{\p F}{\p\phi} = F \bar{F}\ ,\quad G = \bar{F}\ .
 \end{equation}
There are three independent solutions for $F$:
 \begin{equation}
  F_1 = \kappa \tan(\kappa \phi + \rho) + \i \kappa\ ,\quad F_2 =
  - \frc{1}{\phi + \mu}\ ,\quad F_3 = 0\ ,\quad \kappa, \rho, \mu
  \in \fieldR\ .
 \end{equation}
The third solution obviously corresponds to the free case, whereas the
second yields the old NLVT multiplet. As was shown in \cite{IS}, the
first solution, while leading to self-consistent superfield constraints,
gives rise to a differential constraint in $x$-space on the 3-form
component field (dual to $H^\mu$ in the free limit) that cannot be
solved identically in terms of a 2-form gauge potential without
violating locality. One therefore does not obtain a VT multiplet in the
usual sense (although there is a dual description in terms of a vector
multiplet which is well-behaved, see \cite{IS} for details).

We will find, however, that also $F_1$ is associated with a proper VT
multiplet; in the construction of our new NLVT multiplet we encounter
essentially the same differential equation \eqref{dF=FF} for a certain
function of $\phi$, and this time only the first solution turns out to
be fully consistent in the sense that it does not induce constraints on
the $x$-dependence of the multiplet components.
\bigskip

\textbf{The New NLVT Multiplet from Bianchi Identities} 
\medskip

In order to construct the new NLVT multiplet off-shell, we return to the
supersymmetry algebra \eqref{free_alg} and gauge the transformation
generated by $\Dz$ by introducing gauge connections $A_A$ and super
gauge-covariant derivatives
 \begin{equation}
  \cD_A = D_A - g A_A \Dz\ .
 \end{equation}
The connection $A_A$ with subscript $A=\mu$ will be identified with the
1-form in the VT multiplet, and $g\equiv g_1$.

We want to include the central charge generator $\Dz$ among the
gauge-covariant operators, so we impose as a first constraint on the
theory
 \begin{equation} \label{Az=0}
  A_z = 0 \quad\Rightarrow\quad \cD_z = \Dz\ .
 \end{equation}
The commutation relations of the super gauge-covariant derivatives
involve in addition to torsion also the curvature tensors,
 \begin{equation} \label{comm}
  \com{\cD_A}{\cD_B} = - T_{AB}{}^C \cD_C - g \cF_{AB} \Dz\ ,
 \end{equation}
which are given by
 \begin{equation} \label{curv}
  \cF_{AB} = \cD_A A_B - (-)^{\epsilon_A\epsilon_B} \cD_B A_A +
  T_{AB}{}^C\! A_C\ .
 \end{equation}
Gauge covariance of $\cD_A T$, where $T$ is some tensor field
transforming under local central charge transformations with
infinitesimal parameter $C(x)$ as
 \begin{equation}
  \de_C T = g\, C \Dz T\ ,
 \end{equation}
requires $A_A$ to transform according to
 \begin{equation} \label{delA}
  \de_C A_A = D_A C + g\, C \Dz A_A = (D_A + g \cF_{zA}) C\ .
 \end{equation}
In the last step we have used \eqref{Az=0} and $T_{zA}{}^B=0$ in the
equation for $\cF_{zA}$.

The Jacobi identities satisfied by the commutators \eqref{comm} imply
the Bianchi identities (BIs) for the curvatures:
 \begin{equation}
  (\cD_A + g \cF_{zA}) \cF_{BC} + T_{AB}{}^D \cF_{DC} + \text{graded
  cyclic} = 0\ .
 \end{equation}
These equations become nontrivial once we impose constraints on the
curvature components. In order to obtain the new NLVT multiplet, we
first of all adopt the natural constraint
 \begin{equation} \label{con1}
  \cF_\al^i{}^{}_{\,\ad j} = 0\ .
 \end{equation}
With this choice, the BIs read explicitly:
 \begin{align}
  (a) \quad 0 & = (\cD_\al^i + g \cF^{}_z{}_\al^i)\, (T_\be^j{}_\ga^k
	{}^z + g \cF_\be^j{}_\ga^k) + (\cD_\be^j + g \cF^{}_z{}_\be^j)\,
	(T_\ga^k{}_\al^i{}^z + g \cF_\ga^k{}_\al^i) \notag \\*
	& \tab + (\cD_\ga^k + g \cF^{}_z{}_\ga^k)\, (T_\al^i{}_\be^j
	{}^z + g \cF_\al^i{}_\be^j) \notag \\[2pt]
  (b) \quad 0 & = (\bcD_{\ad i} + g \cF_{z\ad i})\, (T_\be^j{}_\ga^k
	{}^z + g \cF_\be^j{}_\ga^k) - \i g\, \de_i^j \si^\mu_{\be\ad}
	\cF_\ga^k{}^{}_\mu - \i g\, \de_i^k \si^\mu_{\ga\ad} \cF_\be^j
	{}^{}_\mu \notag \\[2pt]
  (c) \quad 0 & = (\cD_\mu + g \cF_{z\mu})\, (T_\al^i{}_\be^j{}^z + g
	\cF_\al^i{}_\be^j) + g (\cD_\al^i + g \cF^{}_z{}_\al^i)
	\cF_\be^j{}^{}_\mu + g (\cD_\be^j + g \cF^{}_z{}_\be^j)
	\cF_\al^i{}^{}_\mu \notag \\[2pt]
  (d) \quad 0 & = (\cD_\al^i + g \cF^{}_z{}_\al^i) \cF_{\ad j\mu} +
	(\bcD_{\ad j} + g \cF_{z\ad j}) \cF_\al^i{}^{}_\mu - \i \de^i_j
	\si^\nu_{\al\ad} \cF_{\mu\nu} \notag \\[2pt]
  (e) \quad 0 & = \Dz \cF_\al^i{}_\be^j + \cD_\al^i \cF_\be^j{}^{}_z
	+ \cD_\be^j \cF_\al^i{}^{}_z \notag \\[2pt]
  (f) \quad 0 & = \bcD_{\ad i} \cF_\al^j{}^{}_z + \cD_\al^j \cF_{\ad
	iz} + \i \de_i^j \si^\mu_{\al\ad} \cF_{\mu z} \notag \\[2pt]
  (g) \quad 0 & = \Dz \cF_\al^i{}^{}_\mu + \cD_\al^i \cF_{\mu z} -
	\cD_\mu \cF_\al^i{}^{}_z \notag \\[2pt]
  (h) \quad 0 & = (\cD_\al^i + g \cF^{}_z{}_\al^i) \cF_{\mu\nu} -
	(\cD_\mu + g \cF_{z\mu}) \cF_\al^i{}^{}_\nu + (\cD_\nu + g
	\cF_{z\nu}) \cF_\al^i{}^{}_\mu \notag \\[2pt]
  (i) \quad 0 & = \Dz \cF_{\mu\nu} + \cD_\mu \cF_{\nu z} - \cD_\nu
	\cF_{\mu z} \notag \\*[2pt]
  (j) \quad 0 & = (\cD_\mu + g \cF_{z\mu}) \cF_{\nu\rho} + (\cD_\nu +
	g \cF_{z\nu}) \cF_{\rho\mu} + (\cD_\rho + g \cF_{z\rho})
	\cF_{\mu\nu}\ . \label{BIs}
 \end{align}
All other identities are either satisfied automatically or can be
obtained from the ones above by complex conjugation. The first three
equations have been multiplied with $g$ in order to write them in a
more convenient fashion, using the fact that the torsion components
are constant.

As our second constraint we take\footnote{The constraints \eqref{con1}
and \eqref{con2b} are `representation preserving', in the sense that
one may consistently describe hypermultiplets by the usual
covariantized constraints $\cD_\al^{(i}\varphi^{j)}=0=\bcD_\ad^{(i}
\varphi^{j)}$.}
 \begin{equation} \label{con2b}
  \cF_\al^i{}_\be^j + \cF_\be^i{}_\al^j = 0 \quad\Rightarrow\quad
  T_\al^i{}_\be^j{}^z + g \cF_\al^i{}_\be^j = - \i \ep_{\al\be}
  \ep^{ij} Z(g)\ ,
 \end{equation}
where $Z(g)$ is some unconstrained complex scalar field with $Z(0)=1$.
The BI $(a)$ is then satisfied if
 \begin{equation}
  \cD_\al^i \ln(Z) = g \cF_\al^i{}^{}_z\ .
 \end{equation}
Next we consider the BI $(b)$. Using the result for $\cF_\al^i{}^{}_z$,
we can express the curvature components $\cF_\al^i{}^{}_\mu$ in terms
of $Z$ and its complex conjugate,
 \begin{equation}
  \cF_\al^i{}^{}_\mu = \frc{Z}{2g} (\si_\mu \bcD^i)_\al \ln(Z/
  \bar{Z})\ .
 \end{equation}
The VT multiplet contains a real scalar, therefore one of the two
degrees of freedom contained in (the lowest component of) $Z$ must be
eliminated. We cannot simply take $Z$ to be real or imaginary, because
in either case $\cF_\al^i{}^{}_\mu$ would vanish, which leads to a
trivial supersymmetry transformation of $A_\mu$ (according to
\eqref{curv} $\cF_\al^i{}^{}_\mu=0$ would imply $\cD_\al^i A_\mu=
\cD_\mu A_\al^i$, which reduces to a gauge transformation in the
limit $g=0$). Instead, we require $\ln(Z)$ to be purely imaginary,
 \begin{equation} \label{con2}
  Z = \e^{\i g\phi} \quad\Leftrightarrow\quad \cF_\al^i{}_\be^j =
  \frc{\i}{\raisebox{2pt}{$g$}}\, \ep_{\al\be} \ep^{ij} (1 - \e^{\i
  g\phi})\ ,
 \end{equation}
with $\phi$ a real scalar field that is as yet unconstrained. The
solutions to the BIs of dimension $3/2$ then read
 \begin{equation}
  \cF_\al^i{}^{}_z = \i\, \cD_\al^i \phi\ , \quad \cF_\al^i{}^{}_\mu =
  \i\, \e^{\i g\phi} (\si_\mu \bcD^i \phi)_\al\ .
 \end{equation}

We now solve the remaining Bianchi identities subject to the constraint
\eqref{con2}. Let us consider the BIs of dimension 2 first. $(c)$
implies
 \begin{equation}
  \cF_{\mu z} = \frc{1}{4}\, \bsi^{\ad\al}_\mu \com{\cD_\al^i}{
  \bcD_{\ad i}}\, \phi\ ,
 \end{equation}
and imposes a constraint on $\phi$:
 \begin{equation} \label{con_phi}
  \cD_\al^{(i}\, \bcD_\ad^{j)} \phi = 0\ .
 \end{equation}
The identity $(d)$ gives the remaining field strength,
 \begin{equation}
  \cF_{\mu\nu} = \frc{\i}{4g}\, \big( \e^{\i g\phi}\, \cD^i \si_{\mu\nu}
  \cD_i \e^{-2\i g\phi} - \e^{-\i g\phi}\, \bcD^i \bsi_{\mu\nu} \bcD_i
  \e^{2\i g\phi} \big)\ ,
 \end{equation}
as well as the reality condition
 \begin{equation} \label{real}
  \e^{\i g\phi}\, \cD^{(i} \cD^{j)} \e^{-2\i g\phi} + \e^{-\i g\phi}\,
  \bcD^{(i} \bcD^{j)} \e^{2\i g\phi} = 0\ .
 \end{equation}
With the above results, the BIs $(e)$ and $(f)$ are satisfied
identically. Finally, it is an exercise in index gymnastics to show
that the BIs $(g)$ to $(j)$ of dimension $5/2$ and 3 are consequences
of the other identities and the commutation relations and do not give
any further restrictions.

This completes the evaluation of the Bianchi identities. As a result of
the constraints \eqref{con1} and $\eqref{con2}$ all field strengths can
be expressed in terms of a single real field $\phi$ and its
supercovariant derivatives. However, the Bianchi identities do not
guarantee that the supersymmetry algebra is realized on the whole
$\phi$-multiplet. There arise additional constraints from imposing
the commutation relations \eqref{comm} on the higher component fields.
In order to discuss these constraints and to make contact with the
VT multiplet, we introduce the following notation/identifications:
 \begin{equation}
  \cF_\al^i{}^{}_z = \i \psi_\al^i\ ,\quad \cF_{z\mu} = V_\mu\ ,\quad
  \Dz \phi = U\ .
 \end{equation}
The commutation relations then read
 \begin{align}
  \aco{\cD_\al^i}{\bcD_{\ad j}} & = - \i\, \de^i_j \si^\mu_{\al\ad}
	\cD_\mu & \aco{\cD_\al^i}{\cD_\be^j} & = \i\, \ep_{\al\be}
	\ep^{ij} \e^{\i g\phi} \Dz \notag \\[2pt]
  \com{\cD_\mu}{\cD_\al^i} & = \i g\, \e^{\i g\phi} (\si_\mu \bpsi^i
	)_\al\, \Dz & \com{\Dz}{\cD_\al^i} & = \i g\, \psi_\al^i\, \Dz
	\notag \\[2pt]
  \com{\cD_\mu}{\cD_\nu} & = - g \cF_{\mu\nu}\, \Dz & \com{\Dz}{\cD_\mu}
	& = - g V_\mu\, \Dz\ , \label{comrel}
 \end{align}
where the (anti-) commutators not listed can be obtained by complex
conjugation of the ones above. Furthermore we write
 \begin{equation} \label{Mdef}
  \cD^{(i} \cD^{j)} \phi = 2 g M^{ij}\ .
 \end{equation}
The right-hand side has to be proportional to $g$ since it vanishes for
the free VT multiplet. According to the Bianchi identities, the fields
 \begin{equation}
  \phi\, ,\ \psi_\al^i\, ,\ V_\mu\, ,\ \cF_{\mu\nu}\, ,\ U
 \end{equation}
comprise the covariant components of the supersymmetry multiplet. The
complex SU(2) triplet $M^{ij}$ has mass dimension 3 and can only be a
composite field to be identified later on. The action of the
supersymmetry generators on $V_\mu$ and $\cF_{\mu\nu}$ can be read
directly off the Bianchi identities $(g)$ and $(h)$ respectively,
 \begin{align}
  \cD_\al^i V_\mu & = - \i\, \cD_\mu \psi_\al^i + \i \Dz (\e^{\i g\phi}
	\si_\mu \bpsi^i)_\al \notag \\[2pt]
  \cD_\al^i \cF_{\mu\nu} & = 2\i\, (\cD_{[\mu} + g V_{[\mu}) (\e^{\i
	g\phi} \si_{\nu]} \bpsi^i)_\al + \i g\, \psi_\al^i \cF_{\mu
	\nu}\, , \label{DVDF}
 \end{align}
while for the other fields we find
 \begin{align}
  \cD_\al^i \phi & = \psi_\al^i \notag \\[2pt]
  \cD_\al^i \psi^{\be j} & = \half\, \ep^{ij} \e^{\i g\phi}
	(\cF_{\mu\nu} \si^{\mu\nu} - \i U)_\al{}^\be - \i g\, \ep^{ij}
	\psi_\al^k \psi^\be_k - g\, \de_\al^\be M^{ij} \notag \\[2pt]
  \cD_\al^i \bpsi_\ad^j & = \half\, \ep^{ij} \si^\mu_{\al\ad} (V_\mu
  + \i \cD_\mu \phi) \notag \\[2pt]
  \cD_\al^i U & = \Dz \psi_\al^i - \i g\, \psi_\al^i U\ .
 \end{align}
Here the $\Dz$-transform of $\psi_\al^i$ enters. Using \eqref{con_phi},
we obtain for the action of $\Dz$ on $\cD_\al^i\phi$
 \begin{equation}
  \Dz \cD_\al^i \phi = \e^{\i g\phi}\, \cD_{\al\ad} \bcD^{\ad i} \phi
  + 2\i g\, \Dz \phi\, \cD_\al^i \phi + \tfrac{\i}{3}\, \e^{\i g\phi}\,
  \cD_{\al j} \bcD^{(i} \bcD^{j)} \phi\ ,
 \end{equation}
or, using the notation introduced above and $\bar{M}^{ij}=(M_{ij})^*$,
 \begin{equation}
  \Dz \psi_\al^i = \e^{i g\phi} \big( \si^\mu \cD_\mu \bpsi^i +
  \tfrac{2}{3}\, \i g\, \cD_j \bar{M}^{ij} \big)_\al + 2\i g\,
  \psi_\al^i U\ .
 \end{equation}
So we have obtained a deformation of the VT multiplet provided $V_\mu$
can be identified with the Hodge-dual field strength of a 2-form gauge
potential (to lowest order in $g$). In order to establish the
corresponding constraint on $V_\mu$ we first of all note that
\eqref{Az=0} and \eqref{curv} imply
 \begin{equation} \label{DzA}
  \Dz A_\mu = V_\mu\ .
 \end{equation}
Thus the field strength $\cF_{\mu\nu}$ can be written as
 \begin{equation}
  \cF_{\mu\nu} = (\p_\mu + g V_\mu) A_\nu - (\p_\nu + g V_\nu) A_\mu\ .
 \end{equation}
It follows that
 \begin{equation}
  \Dz \cF_{\mu\nu} = \cD_\mu V_\nu - \cD_\nu V_\mu\ ,
 \end{equation}
in accordance with the BI $(i)$.

Now, the result for $\cF_\al^i{}^{}_z$ requires that the operator
$\e^{\i g\phi}\Dz$ commute with $\cD_\al^i$. When applied to $\cD_\al^i
\phi$ and its complex conjugate, two nontrivial identities follow from
the parts antisymmetric in the SU(2) indices:
 \begin{align}
  0 & = \half\, \e^{-2\i g\phi} \com{\e^{\i g\phi} \Dz}{\cD^{\al i}}\,
	\cD_{\al i} \phi \notag \\*[2pt]
  & = \i\, (\cD_\mu \cD^\mu \phi - \Dz^2 \phi) - \quart\, \cD^{\ad\al}
	\com{\cD_\al^i}{\bcD_{\ad i}} \phi - g (\Dz \phi)^2 + 2\i g\,
	\e^{\i g\phi}\, \bcD_i \phi\, \Dz \bcD^i \phi \notag \\
  & \tab + \tfrac{\i}{6}\, \cD_i \cD_j\, \bcD^{(i} \bcD^{j)} \phi
	\\[6pt]
  0 & = \i\, \e^{-\i g\phi} \com{\e^{\i g\phi} \Dz}{\cD_\al^i}\,
	\bcD_{\ad i} \phi \notag \\[2pt]
  & = - \tfrac{\i}{2}\, \e^{-\i g\phi}\, \cD_{\be\ad} \com{\cD_{\al
	i}}{\cD^{\be i}} \phi + g\, \Dz \phi\, (\i\, \cD_{\al\ad} \phi
	- \half \com{\cD_\al^i}{\bcD_{\ad i}} \phi) \notag \\[2pt]
  & \tab + \tfrac{\i}{2} \Dz \com{\cD_\al^i}{\bcD_{\ad i}} \phi + 4
	g\, \bcD_{\ad i} \phi\, \Dz \cD_\al^i \phi + \tfrac{1}{3}\,
	\e^{-\i g\phi}\, \cD_{\al i} \bcD_{\ad j}\, \cD^{(i} \cD^{j)}
	\phi\ .
 \end{align}
The imaginary parts of these equations give the action of $\Dz$ on $U$
and $V_\mu$ respectively,
 \begin{align}
  \Dz U & = \cD_\mu \cD^\mu \phi + g\, \big[ \e^{-\i g\phi} \psi_i
	\Dz \psi^i + \tfrac{1}{6}\, \cD_i \cD_j \bar{M}^{ij} +
	\text{c.c.} \big] \\[2pt]
  \Dz V^\mu & = - \cD_\nu (\cF^{\mu\nu} + g \Sigma^{\mu\nu}) - g
	(\tilde{\cF}^{\mu\nu} + g \tilde{\Sigma}^{\mu\nu}) \cD_\nu
	\phi + g U \cD^\mu \phi \notag \\
  & \tab - \i g\, \big[ \psi^i \si^\mu \Dz \bpsi_i + \tfrac{1}{6}\,
	\e^{-\i g\phi} \cD_i \si^\mu \bcD_j M^{ij} - \text{c.c.}
	\big]\ , \label{DzV}
 \end{align}
whereas the real parts provide the Bianchi-like constraints
 \begin{align}
  \cD_\mu V^\mu & = g\, U^2 + \i g\, \big[ \e^{-\i g\phi} \psi_i \Dz
	\psi^i - \tfrac{1}{6}\, \cD_i \cD_j \bar{M}^{ij} - \text{c.c.}
	\big] \label{DV} \\[2pt]
  \cD_\nu (\tilde{\cF}^{\mu\nu} + g \tilde{\Sigma}^{\mu\nu}) & = g
	(\cF^{\mu\nu} + g \Sigma^{\mu\nu}) \cD_\nu \phi + g\,
	U V^\mu - g \Dz \Lambda^\mu \notag \\*
  & \tab + \tfrac{1}{6} g\, \big[ \e^{-\i g\phi} \cD_i \si^\mu \bcD_j
	M^{ij} + \text{c.c.} \big]\ . \label{DF}
 \end{align}
Here we have introduced the following composite objects:
 \begin{equation}
  \Lambda^\mu = \psi^i \si^\mu \bpsi_i\ ,\quad \Sigma^{\mu\nu} = \i
  (\e^{-\i g\phi} \psi^i \si^{\mu\nu} \psi_i - \e^{\i g\phi} \bpsi^i
  \bsi^{\mu\nu} \bpsi_i)\ .
 \end{equation}
Clearly, the equations \eqref{DV} and \eqref{DF} can be solved only
for an appropriate choice of $M^{ij}$.

Furthermore, we have a consistency condition which follows from the
fact that the spinor derivatives $\cD_\al^i$ anticommute when
symmetrized in the SU(2) indices:
 \begin{equation} \label{M1}
  \cD_\al^{(i} M^{jk)} = 0\ .
 \end{equation}
Finally, we have to take into account the reality condition
\eqref{real}, which when written in terms of $M^{ij}$ reads
 \begin{equation} \label{M2}
  \e^{-\i g\phi} (M^{ij} - \i \psi^i \psi^j) = \e^{\i g\phi}
  (\bar{M}^{ij} + \i \bpsi^i \bpsi^j)\ .
 \end{equation}
Note that \eqref{M1} and \eqref{M2} imply $\cD_\al^{(i}\bar{M}^{jk)}
=0$.

We must determine $M^{ij}$ such that the constraints are satisfied
identically. Toward this end, we make an ansatz for $M^{ij}$:
 \begin{equation}
  M^{ij} = F(g\phi) \psi^i \psi^j + G(g\phi) \bpsi^i \bpsi^j\ ,
 \end{equation}
where $F$ and $G$ are arbitrary complex functions of the dimensionless
argument $g\phi$. The constraints \eqref{M1} and \eqref{M2} then
require
 \begin{equation}
  G' = 2 FG\ ,\quad G = \e^{2\i g\phi} (\bar{F} + \i) \ ,
 \end{equation}
where a prime denotes differentiation with respect to $g\phi$.
Introducing $H=F-\i$, we obtain
 \begin{equation} \label{H}
  H' = 2 H \bar{H}\ .
 \end{equation}
Now, this is the same differential equation \eqref{dF=FF} that we
encountered in the construction of the old NLVT multiplet, with the
crucial difference that there it was $F$ itself which had to satisfy
\eqref{H}. We can copy the solutions, which are now shifted by the
imaginary unit,
 \begin{equation}
  F_1 = \kappa \tan(2\kappa g \phi + \rho) + \i (\kappa + 1)\ ,\quad
  F_2 = \i - \frc{1}{2g\phi + \mu}\ ,\quad F_3 = \i\ ,\quad \kappa,
  \rho, \mu \in \fieldR\ .
 \end{equation}
We now have to check which of the three possible $M^{ij}$ is
compatible with the remaining constraints, thereby giving a consistent
deformation.

Let us first consider \eqref{DV}. A necessary condition for this
constraint to admit a local solution is that all terms involving
$U^2$ cancel. With the ansatz for $M^{ij}$ inserted, we find
 \begin{equation}
  \cD_\mu V^\mu = g \im F\, U^2 + \dots\ ,
 \end{equation}
where the terms omitted are at most linear in $U$. This implies that
$F$ must be real, which singles out the first solution $F_1$ with
$\kappa=-1$. The remaining parameter $\rho$ can be absorbed by a
redefinition
 \begin{equation}
  \phi \rightarrow \phi + \rho / 2g\ ,\quad \psi_\al^i \rightarrow
  \e^{\i\rho/4}\, \psi_\al^i\ ,\quad \cD_\al^i \rightarrow
  \e^{\i\rho/4}\, \cD_\al^i
 \end{equation}
(with all other bosonic fields being invariant), which is an
automorphism of the supersymmetry algebra \eqref{comrel}. So without
loss of generality the only possible candidate for $M^{ij}$ is given by
 \begin{equation} \label{Mij}
  M^{ij} = \tan(2g\phi)\, \psi^i \psi^j + \frc{\i}{\cos(2g\phi)}\,
  \bpsi^i \bpsi^j\ .
 \end{equation}
Using this expression in \eqref{DV} and \eqref{DzV}, we find after some
algebra
 \begin{equation} \label{DW}
  \cD_\mu W^\mu = \half g\, \cF_{\mu\nu} G^{\mu\nu}\ ,\quad \Dz W^\mu
  = - (\cD_\nu - g V_\nu) G^{\mu\nu}\ ,
 \end{equation}
where
 \begin{align}
  W^\mu & = \cos(2g\phi)\, V^\mu + 2g \sin(2g\phi)\, \Lambda^\mu
	\notag \\[2pt]
  G_{\mu\nu} & = \cos(2g\phi)\, (\cF_{\mu\nu} + 2g \Sigma_{\mu\nu})
	+ \sin(2g\phi)\, (\tilde{\cF}_{\mu\nu} + 2g \tilde{\Sigma}_{
	\mu\nu})\ .
 \end{align}
\eqref{DW} can easily be solved in terms of an unconstrained 2-form
gauge potential $B_{\mu\nu}$,
 \begin{equation} \label{W}
  W^\mu = \half \ep^{\mu\nu\rho\si} \cD_\nu B_{\rho\si}\ ,
 \end{equation}
with a covariant central charge transformation generated by
 \begin{equation} \label{DzB}
  \Dz B_{\mu\nu} = \half \ep_{\mu\nu\rho\si} G^{\rho\si}\ .
 \end{equation}
The field strength that appears in the commutation relations, the
symmetry transformations, and finally in the action, is $V_\mu$ rather
than $W_\mu$. By splitting $\cF_{\mu\nu}=F_{\mu\nu}+2gV_{[\mu}A_{\nu]}$
we can bring \eqref{W} into the form
 \begin{align}
  K^{\mu\nu} V_\nu = \mathcal{H}^\mu & \equiv H^\mu - 2g \sin(2g\phi)\,
	\Lambda^\mu + g \cos(2g\phi)\, (F^{\mu\nu} + 2g \Sigma^{\mu\nu})
	A_\nu \notag \\
  & \tab + g \sin(2g\phi)\, (\tilde{F}^{\mu\nu} + 2g \tilde{\Sigma}^{
	\mu\nu}) A_\nu\ , \label{calH}
 \end{align}
where the matrix $K^{\mu\nu}$ is given by
 \begin{equation}
  K^{\mu\nu} = \cos(2g\phi)\, \big[ \eta^{\mu\nu} (1 - g^2 A^\rho\!
  A_\rho) + g^2 A^\mu A^\nu \big]\ .
 \end{equation}
The nonpolynomial structure of the model now arises by inverting
$K^{\mu\nu}$:
 \begin{equation} \label{VKH}
  V_\mu = (K^{-1})_{\mu\nu}\, \mathcal{H}^\nu\ ,\quad (K^{-1})_{\mu\nu}
  = \frc{\eta_{\mu\nu} - g^2 A_\mu A_\nu}{\cos(2g\phi)\, (1 - g^2
  A^\rho\! A_\rho)}\ .
 \end{equation}
Note that in the on-shell version of the new NLVT multiplet given in
\cite{T} the nonpolynomial dependence on $gA_\mu$ could be avoided by
means of a first-order formulation (where ``first-order'' now refers to
the order of the 2-form in the action rather than that of the coupling
constant) and emerged only after eliminating an auxiliary vector field,
which made the construction possible in the first place. Here it is a
result of the Bianchi identities and present from the outset, without
specifying an invariant action. An off-shell first-order formulation
would require a whole multiplet of auxiliary fields, and it is not clear
to us whether it actually exists.

The last constraint to check is \eqref{DF}. With $M^{ij}$ as in
\eqref{Mij} it turns out to be equivalent to the Bianchi identity
$(j)$, so we have satisfied all constraints and consistency conditions.
\medskip

It remains to determine the supersymmetry transformations of the gauge
potentials $A_\mu$ and $B_{\mu\nu}$. The action of $\cD_\al^i$ on the
former follows directly from the definition of the curvatures in
\eqref{curv},
 \begin{equation}
  \cD_\al^i A_\mu = \cF_\al^i{}^{}_\mu + \cD_\mu A_\al^i =
  \cF_\al^i{}^{}_\mu + g A_\mu \cF_\al^i{}^{}_z + \p_\mu A_\al^i\ .
 \end{equation}
The last term is a gauge transformation, which may be dropped. We then
have
 \begin{equation}
  \cD_\al^i A_\mu = \i\, \e^{\i g\phi} (\si_\mu \bpsi^i)_\al + \i g\,
  \psi_\al^i A_\mu\ ,
 \end{equation}
and the fields $A_\al^i$, $\bar{A}_{\ad i}$ occur nowhere explicitly
anymore. As a direct consequence of the Bianchi identities \eqref{BIs}
we find the following supersymmetry commutators:
 \begin{align}
  \aco{\cD_\al^i}{\bcD_{\ad j}} A_\mu & = - \i \de^i_j \si^\nu_{\al
	\ad} \big[ \p_\nu A_\mu - (\p_\mu + g V_\mu) A_\nu \big]
	\notag \\[2pt]
  \aco{\cD_\al^i}{\cD_\be^j} A_\mu & = \frc{\i}{\raisebox{2pt}{$g$}}\,
	\ep_{\al\be} \ep^{ij}\, (\p_\mu + g V_\mu)\, \e^{\i g\phi}
	\notag \\[2pt]
  \com{\Dz}{\cD_\al^i} A_\mu & = \i (\p_\mu + g V_\mu) \psi_\al^i\ .
 \end{align}
There appear additional gradients $\p_\mu(\dots)$ on the right-hand
sides due to our dropping the $\p_\mu A_\al^i$ term; they combine with
$gV_\mu$ into the full gauged central charge transformation \eqref{delA}
of $A_\mu$.

The supersymmetry transformation of $B_{\mu\nu}$ can most easily be
obtained from the one of its $\Dz$ transform, as the following
calculation shows:
 \begin{align}
  \Dz\, \cD_\al^i B_{\mu\nu} & = \cD_\al^i \tilde{G}_{\mu\nu} +
	\com{\Dz}{\cD_\al^i} B_{\mu\nu} \notag \\
  & = \Dz \big[ 2 \cos(2g\phi)\, \si_{\mu\nu} \psi^i - 2 g\,
	\e^{-\i g\phi} A_{[\mu} \si_{\nu]} \bpsi^i \big]_\al + 2\,
	\p_{[\mu} (\e^{-\i g\phi} \si_{\nu]} \bpsi^i)_\al \notag \\
  & \tab - \i g\, \psi_\al^i \tilde{G}_{\mu\nu} + \com{\Dz}{\cD_\al^i}
	B_{\mu\nu}\ .
 \end{align}
If we set
 \begin{equation}
  \cD_\al^i B_{\mu\nu} = 2 \cos(2g\phi)\, (\si_{\mu\nu} \psi^i)_\al
  - 2 g\, \e^{-\i g\phi} A_{[\mu} (\si_{\nu]} \bpsi^i)_\al\ ,
 \end{equation}
we find a commutator
 \begin{equation}
  \com{\Dz}{\cD_\al^i} B_{\mu\nu} = \i g\, \psi_\al^i \tilde{G}_{\mu\nu}
  - 2\, \p_{[\mu} (\e^{-\i g\phi} \si_{\nu]} \bpsi^i)_\al
 \end{equation}
which involves in addition to the central charge transformation a
standard gauge transformation of $B_{\mu\nu}$ of the form
\eqref{free_del}. For the odd commutators we obtain
 \begin{align}
  \aco{\cD_\al^i}{\bcD_{\ad j}} B_{\mu\nu} & = - \i \de^i_j
	\si^\rho_{\al\ad} \big[ \cD_\rho B_{\mu\nu} + 2\, \p_{[\mu}
	(B_{\nu]\rho} - \eta_{\nu]\rho} \sin(2g\phi) / 2g) \big]
	\notag \\[2pt]
  \aco{\cD_\al^i}{\cD_\be^j} B_{\mu\nu} & = \i \ep_{\al\be} \ep^{ij}\,
	\big[ \e^{\i g\phi} \tilde{G}_{\mu\nu} + 2\i\, \p_{[\mu}
	(\e^{-\i g\phi} A_{\nu]}) \big]\ .
 \end{align}
The algebra of supersymmetry and gauged central charge transformations
as given in equations $(12)$ to $(15)$ of ref.\ \cite{T} now closes
off-shell.
\bigskip

\textbf{The Linear Multiplet and Invariant Actions} 
\medskip

With an off-shell formulation of the new NLVT multiplet at hand, we now
have to determine a supersymmetric and gauge-invariant action. Similarly
to the previously known versions of the VT multiplet, it may be derived
from a linear multiplet. However, since the commutation relations
\eqref{comrel} differ from the usual ones considered in the literature,
the constraints that determine the linear multiplet turn out to require
modification: let us consider fields $L^{ij}$ that satisfy
 \begin{equation} \label{lin_con}
  L^{ij} = L^{ji} = (L_{ij})^*\ ,\quad \cD_\al^{(i} \big(\e^{\i g\phi}
  L^{jk)} \big) = 0\ .
 \end{equation}
The higher components of the multiplet consist of a Weyl spinor doublet,
a complex scalar and a real vector. We define these fields such that
their transformations are as simple as possible:
 \begin{align}
  \la_\al^i & = (\cD_{\al j} + \i g\, \psi_{\al j}) L^{ij} \notag
	\\[2pt]
  S & = \half \cD_i \cD_j L^{ij} + \i g\, \psi_i \cD_j L^{ij} +
	\tfrac{\i}{2}\, g^2 (2 M_{ij} + \i \psi_i \psi_j) L^{ij}
	\notag \\[2pt]
  K^\mu & = \i\, \cD_i \si^\mu \bcD_j L^{ij} - g (\psi_i \si^\mu \bcD_j
	+ \bpsi_i \bsi^\mu \cD_j) L^{ij} + \i g^2\, \psi_i \si^\mu
	\bpsi_j\, L^{ij}\ .
 \end{align}
The determination of the supersymmetry transformations is
straightforward. We find
 \begin{align}
  \cD_\al^i L^{jk} & = \tfrac{2}{3}\, \ep^{i(j} \la_\al^{k)} - \i g\,
	\psi_\al^i L^{jk} \notag \\[2pt]
  \cD_\al^i \la_\be^j & = \half \ep_{\al\be} \big[ \ep^{ij} S - 3
	\i\, \Dz (\e^{\i g\phi} L^{ij}) \big] - \i g\, \psi_\al^i
	\la_\be^j \notag \\[2pt]
  \cD_\al^i \bla_\ad^j & = - \tfrac{\i}{4} \si^\mu_{\al\ad} \big[
	\ep^{ij} K_\mu + 6 (\cD_\mu - g V_\mu) L^{ij} \big] - \i g\,
	\psi_\al^i \bla_\ad^j \notag \\[2pt]
  \cD_\al^i S & = \i \Dz (\e^{\i g\phi} \la_\al^i) - \i g\, \psi_\al^i
	S \notag \\[2pt]
  \cD_\al^i \bar{S} & = - 2\i\, (\cD_\mu - g V_\mu) (\si^\mu \bla^i)_\al
	+ \i \Dz \big( \e^{-\i g\phi} \la^i - 6\i g\, \e^{-\i g\phi}
	\psi_j L^{ij} \big)_\al - \i g\, \psi_\al^i \bar{S} \notag
	\\[2pt]
  \cD_\al^i K^\mu & = 4 (\cD_\nu - g V_\nu) (\si^{\mu\nu} \la^i)_\al -
 	2 \Dz \big( \e^{\i g\phi} \si^\mu \bla^i + 3 \i g\, \e^{\i g
	\phi} \si^\mu \bpsi_j L^{ij} \big)_\al - \i g\, \psi_\al^i
	K^\mu\ .
 \end{align}
As usual, however, the commutation relations hold only if the vector
$K^\mu$ satisfies a constraint. Here we must require
 \begin{align}
  & (\cD_\mu - g V_\mu) K^\mu = \Dz X \notag \\[2pt]
  & X = \e^{-\i g\phi} (\i S - 4 g\, \psi^i \la_i) - \e^{\i g\phi}
	(\i \bar{S} - 4 g\, \bpsi_i \bla^i)  + 6 g^2\, \e^{-\i g\phi}
	(M_{ij} - \i \psi_i \psi_j) L^{ij}\ .
 \end{align}
Note that $X$ is real according to \eqref{M2}. The constraint implies
that a spacetime integral over
 \begin{equation}
  \mathcal{L} = X + g A_\mu K^\mu
 \end{equation}
is invariant under gauged central charge transformations:
 \begin{align}
  \de_C \mathcal{L} & = g C\, (\cD_\mu - g V_\mu) K^\mu + g (\p_\mu C
	+ g V_\mu C) K^\mu + g A_\mu\, g C \Dz K^\mu \notag \\
  & = g\, \p_\mu (C K^\mu)\ .
 \end{align}
A little more effort is needed to show that $\int\!d^4x\,\mathcal{L}$
is also supersymmetric. We find
 \begin{equation}
  \cD_\al^i \mathcal{L} = - 2\, \p_\mu \big( \e^{\i g\phi} \si^\mu
  \bla^i + 3\i g\, \e^{\i g\phi} L^{ij} \si^\mu \bpsi_j + 2 g A_\nu
  \si^{\mu\nu} \la^i \big)_\al\ .
 \end{equation}
Hence, we have an action rule at our disposal: If a composite field
$L^{ij}$ satisfies the constraints \eqref{lin_con}, then the expression
 \begin{align}
  \mathcal{L} & = \frc{1}{12}\, \e^{-\i g\phi} \big( \i\, \cD_i \cD_j
	+ 6 g\, \psi_i \cD_j + 4 g^2 M_{ij} + \i g^2 \psi_i \psi_j
	\big)\, L^{ij} \notag \\
  & \tab + \frc{\i}{12}\, g A_\mu \big( \cD_i \si^\mu \bcD_j + 2\i g\,
	\psi_i \si^\mu \bcD_j + g^2\, \psi_i \si^\mu \bpsi_j \big)\,
	L^{ij} + \text{c.c.} \label{cand_L}
 \end{align}
is a candidate for a Lagrangian.

Let us apply this construction principle to the new NLVT multiplet
itself. We make the following ansatz for $L^{ij}$:
 \begin{equation} \label{ansatz_L}
  L^{ij} = \al(g\phi)\, \psi^i \psi^j + \bar{\al}(g\phi)\, \bpsi^i
  \bpsi^j\ .
 \end{equation}
It already satisfies the symmetry and reality properties, while the
third constraint in \eqref{lin_con} requires furthermore
 \begin{equation} \label{diff_al}
  \cos(2g\phi)\, (\al' - \i \al) + 2\i\, \bar{\al} = 0\ .
 \end{equation}
Modulo real multiplicative constants, the two linearly independent
solutions to this differential equation are given by
 \begin{equation} \label{sol_al}
  \al_1 = \frc{\e^{-\i g\phi}}{\cos(2g\phi)}\ ,\quad \al_2 = 2 g
  \phi\, \alpha_1 - \i\, \e^{\i g\phi}\ .
 \end{equation}
Now, the scalar $\phi$ was defined only modulo $2\pi/g$, see
\eqref{con2}, so one would expect the action to be invariant under
shifts $\phi\rightarrow\phi+2\pi/g$. While $L_1^{ij}$, and thus the
corresponding Lagrangian $\mathcal{L}_1$, is indeed periodic, the
solution $L_2^{ij}$ shifts by $4\pi L_1^{ij}$. Hence, $\mathcal{L}_1$
can only be a total derivative.

The Lagrangian as it follows from \eqref{cand_L} and the ansatz
\eqref{ansatz_L} reads
 \begin{align}
  \mathcal{L} & = - \im (\e^{\i g\phi} \al) \big[\, \half \cD^\mu
	\phi\, \cD_\mu \phi - \half V^\mu V_\mu - \quart \cF^{\mu\nu}
	\cF_{\mu\nu} + \half U^2 - g A_\mu \cF^{\mu\nu} V_\nu \notag \\
  & \mspace{124mu} + g\, U\! A_\mu \cD^\mu \phi + g A_\mu
	\tilde{\cF}^{\mu\nu} \cD_\nu \phi\, \big] \notag \\
  & \tab - \re (\e^{\i g\phi} \al) \big[\, \quart \tilde{\cF}^{\mu
	\nu} \cF_{\mu\nu} + V^\mu \cD_\mu \phi + g\, U\! A_\mu V^\mu
	+ g A_\mu \tilde{\cF}^{\mu\nu} V_\nu \notag \\
  & \mspace{124mu} + g A_\mu \cF^{\mu\nu} \cD_\nu \phi\, \big]
	\notag \\
  & \tab + \text{fermion terms}\ .
 \end{align}
As anticipated, $\alpha_1$ gives rise to a total derivative,
 \begin{equation}
  \mathcal{L}_1 = - \frc{1}{2g}\, \p_\mu \big( \sin(2g\phi)\, V^\mu
  + g\, \tilde{G}^{\mu\nu}\! A_\nu + \text{fermion terms} \big)\ .
 \end{equation}
$\alpha_2$ on the other hand yields the sought nontrivial off-shell
Lagrangian:
 \begin{align}
  \mathcal{L}_2 & = \frc{1}{2} \cos(2g\phi)\, \big[ \cD^\mu \phi\,
	\cD_\mu \phi - V^\mu V_\mu + U^2 + 2 g\, U\! A_\mu \cD^\mu
	\phi \big] \notag \\
  & \tab - \frc{1}{4}\, G^{\mu\nu} (\cF_{\mu\nu} + 4 g A_\mu V_\nu)
	- \p_\mu \big( \sin(2g\phi)\, \phi V^\mu + g \phi\,
	\tilde{G}^{\mu\nu}\! A_\nu \big) \notag \\
  & \tab + \text{fermion terms}\ .
 \end{align}
If we insert the explicit expressions for the composite fields
$G_{\mu\nu}$ and $\cF_{\mu\nu}$, the $V_\mu$-dependent terms combine
into $-\half V^\mu K_{\mu\nu}V^\nu$. Using the solution \eqref{VKH}
to the Bianchi identities, we eventually find (dropping the total
derivative)
 \begin{align}
  \mathcal{L}_2 & = - \frc{1}{2}\, \mathcal{H}^\mu (K^{-1})_{\mu\nu}
	\mathcal{H}^\nu - \frc{1}{4} \cos(2g\phi)\, F^{\mu\nu}
	F_{\mu\nu} - \frc{1}{4} \sin(2g\phi)\, \tilde{F}^{\mu\nu}
	F_{\mu\nu} \notag \\
  & \tab + \frc{1}{2} \cos(2g\phi)\, \p^\mu \phi\, \p_\mu \phi +
	\frc{1}{2} \cos(2g\phi)\, (1 - g^2 A^\mu\! A_\mu)\, U^2
	\notag \\
  & \tab - \i \cos(2g\phi)\, \big( \psi^i \si^\mu \overset{
	\leftrightarrow}{\p_\mu} \bpsi_i \big) - g F_{\mu\nu} \big(
	\cos(2g\phi) \Sigma^{\mu\nu} + \sin(2g\phi) \tilde{\Sigma}^{
	\mu\nu} \big) \notag \\
  & \tab - \frc{g^2}{2\cos(2g\phi)}\, \big( \Sigma^{\mu\nu} \Sigma_{
	\mu\nu} + 2 \Lambda^\mu \Lambda_\mu \big)\ .
 \end{align}
This Lagrangian differs from the one found in \cite{T} in the terms
involving the auxiliary field $U$. Due to the presence of the inverse
matrix $K^{-1}$, the action is nonpolynomial in the combination
$gA_\mu$, in addition to the less unusual nonpolynomial dependence on
the scalar. Since $K^{-1}$ contains no derivatives, however, the
action is in fact local.
\bigskip

\textbf{Conclusions} 
\medskip

We have given a detailed derivation of the off-shell transformation
rules and action for our new nonlinear vector-tensor multiplet and
revealed its geometric origin. Starting from a set of constraints on
the field strengths associated with a local central charge in the $N=2$
supersymmetry algebra, we have shown how the Bianchi identities give
rise to the field content of a VT multiplet. Accordingly, its covariant
components occur as field strengths in the supersymmetry commutation
relations, a novel feature not observed before. What is more, some of
these field strengths turned out to depend nonpolynomially on the vector
gauge field.

We then introduced a new coupling of the linear multiplet to the VT
multiplet, which provided us with an action formula similar to those
used in the construction of models with central charges gauged by a
vector or supergravity multiplet. From this formula we subsequently
derived the invariant action for the new NLVT multiplet itself.
For central charges gauged by a vector multiplet the action rule takes
a particularly simple form in harmonic superspace \cite{DIKST}. It would
be interesting if the results presented here can also be formulated in
harmonic superspace.

An unusual feature of our model is that central charge and supersymmetry
transformations do not commute. While for the previous versions of the
VT multiplet with local central charge the commutator vanishes modulo
ordinary gauge transformations of the vector and tensor, here it
involves another central charge transformation. The reason is that by
gauging the central charge the corresponding transformation of the
vector is identified with its gauge transformation. A similar
phenomenon, a nonvanishing commutator of supersymmetry and gauge
transformations, was observed in a large family of four-dimensional
$N=1$ supersymmetric gauge theories involving 1-forms and 2-forms
constructed in \cite{BT}. And indeed we showed in \cite{T} that the new
NLVT multiplet is but a special member of this family with a second
supersymmetry.

At last we remark that in the case of local supersymmetry central
charges necessarily have to be gauged, so our efforts may be viewed as
preparatory steps toward a coupling of the VT multiplet to supergravity.
For the linear and old nonlinear VT multipet couplings to superconformal
gravity have already been constructed in \cite{CdWFKST2}, where again it
was necessary to include a vector multiplet that provides the gauge
field for local central charge transformations. Upon reduction to
Poincar\'e supergravity this gauge field is identified with the
graviphoton. This leads us to expect a fundamentally different
supergravity coupling of the new NLVT multiplet, since the multiplet
gauges its central charge by itself and therefore is likely to be
invariant under the gauge transformation associated with the
graviphoton. An answer to the question whether the new NLVT multiplet
can be coupled to supergravity at all, however, requires further study.
\bigskip

\textbf{Acknowledgements} 
\medskip

I would like to thank Friedemann Brandt and Sergei Kuzenko for helpful
discussions. This work was supported by the Deutsche
Forschungsgemeinschaft.

\end{document}